\documentclass[a4paper]{jpconf}
\usepackage{graphicx}
\begin{document}
\title{An analytic relation between the fractional parameter in the Mittag--Leffler function 
and the chemical potential in the  Bose--Einstein distribution through
 the analysis of the NASA COBE monopole data}
\author{Minoru Biyajima$^1$, Takuya Mizoguchi$^2$ and Naomichi Suzuki$^3$ }
 \address{$^1$ Department of Physics, Shinshu University, Matsumoto 390-8621, Japan }
 \address{$^2$ National Institute of Technology, Toba College, Toba 517-8501, Japan}
 \address{$^3$ Matsumoto University, Matsumoto, 390-1295, Japan}
\ead{suzuki@matsu.ac.jp}

\begin{abstract}
To extend the Bose-Einstein (BE) distribution to fractional order, we turn our attention to the differential 
equation, $df/dx =-f-f^2$. It is satisfied with the stationary solution,  $f(x)=1/(e^{x+\mu}-1)$, of the Kompaneets equation, 
where $\mu$ is the  constant chemical potential. 
Setting $R=1/f$, we obtain a linear differential equation for $R$. 
Then, the Caputo fractional derivative of order  $p$ ($p>0$) is introduced in place of the derivative of $x$,  
and fractional BE distribution is obtained, where function  ${\rm e}^x$ is replaced by 
 the Mittag--Leffler (ML) function $E_p(x^p)$.  
Using the integral representation of the ML function, we obtain a new formula. 
Based on the analysis of the NASA COBE monopole data, an identity $p\simeq e^{-\mu}$ is found.
\end{abstract}

\section{\label{sec1}Introduction}
The COBE FIRAS experiments have shown that the cosmic microwave background (CMB) radiation 
 spectrum is well described by the Planck distribution with temperature, $T = 2725.0 \pm 1$ mK
~\cite{Mather1994, COBE2005}. 
 Furthermore, a slight distortion from the Planck distribution in the 
photon number distribution, $f(x)$, is observed. 
It is expressed by
 \begin{eqnarray} 
     f(x) =  1/(e^{x + \mu} - 1 ),  \quad   x = h\nu/(kT),  \label{eq.int01} 
 \end{eqnarray}
where $\mu$ is the dimensionless constant chemical potential and $\nu$ is the frequency of photon. 
The measured value is $\mu=(-1 \pm 4)\times 10^{-5}$ or $|\mu|<9\times 10^{-5}$ 
with 95\% confidence~\cite{Mather1994, COBE2005}.
  
Equation (\ref{eq.int01}) is known as a stationary solution of the 
Kompaneets equation\cite{Komp1957}:
 \begin{eqnarray} 
   \frac{\partial f}{\partial t} =  \frac{k T_e}{m_e c^2}\,\frac{n_e\sigma_e}{c}\,{x_e}^{-2}
   \frac{\partial }{\partial x_e}\,{x_e}^4 \Bigl( {\partial f}/{\partial x_e} + f + f^2 \Bigr),
            \label{eq.int02} 
 \end{eqnarray}
where $n_e$ is the electron density, $\sigma_e$ is 
the Thomson scattering cross-section and $x_e=h\nu/(kT_e)$.  Equation (\ref{eq.int02}) describes 
the photon distribution, which obeys the Planck distribution at the initial stage, and is affected by the elastic 
$e$-$\gamma$ scatterings in the expanding universe.  

In the theories of stochastic processes, in order to take a sort of memory effect into account, 
fractional calculus is introduced~\cite{Caputo1967,Podlubny1999,Metzler2000,West2003}.
Based on the Caputo derivative~\cite{Caputo1967}, Ertik et al. proposed a generalized 
BE distribution~\cite{Ertik2009},  $f(x) = 1/( E_p(x)  - 1 )$, 
where $E_p(x)$ denotes the Mittag-Leffler (ML) function defined by 
 \begin{eqnarray}
     E_{p}(x) = \sum_{n=0}^\infty {x^n}/{\Gamma(np + 1)}, \quad p>0.  \label{eq.int04}
 \end{eqnarray}
To extend the BE distribution to fractional order, we turn our attention to the equation, 
 \begin{eqnarray}
    {df(x)}/{dx} = - af(x) - b{f(x)}^2,   \label{eq.int05}
 \end{eqnarray}
where $a$  and $b$ are  constant. 
If $a=b=1$, Eq.(\ref{eq.int05}) reduces to the equation which is adopted by 
Planck~\cite{Planck1900b,Tsallis2009} to derive the blackbody radiation law, 
and is satisfied with the stationary solution (\ref{eq.int01})  of the Kompaneets equation (\ref{eq.int02}).
Putting $f=1/R$, we obtain the linear differential equation, 
 \begin{eqnarray}
       {dR}/{dx} = a R + b.   \label{eq.int06}
 \end{eqnarray}
In \ref{apdx1}, the Caputo fractional derivative is introduced  into Eq.~(\ref{eq.int06}) in place of 
the derivative $x$, and a fractional BE and other distributions are obtained.  

In  \cite{Biya2015}, we have applied the Riemann--Liouville fractional derivative to 
obtain a fractional BE distribution $f(x)=1/(E_p(x^p)-1)$, and we have investigated the NASA COBE monopole data 
using BE and fractional BE distributions. 
The photon spectrum given from Eq.(\ref{eq.int01}) is written as
\begin{eqnarray}
     U^{\rm BE}(x,\mu) = C_B/( e^{x+\mu} - 1 ),    \label{eq.int07}
\end{eqnarray}
where $x=h\nu/(kT)$ and  $C_B=2h\nu^3/c^2$. 
On the other hand, the photon spectrum in the Universe, based on the fractional calculus, is given by
\begin{eqnarray}
     U(x,p) = C_B/( E_p(x^p)  - 1 ).    \label{eq.int08}
\end{eqnarray}

 From the analysis of NASA COBE monopole data~\cite{COBE2005} ,  
the following values of parameters  are estimated \cite{Biya2015}:   from Eq.(\ref{eq.int07}), $T=2.72501 \pm 0.00002$ K and  
$\mu=(-1.1\pm 3.2) \times 10^{-5}$, and   from Eq.(\ref{eq.int08}), $T=2.72501 \pm 0.00003$ K and $p-1 = 
(1.1\pm 3.5) \times 10^{-5}$.  
Then, we estimated a relation between $\mu$ and $p$ as  $\mu \approx 1 - p$.
  
In the present study, the COBE monopole data~\cite{COBE2005} is analyzed by the use of an integral 
 representation of the ML function~\cite{Podlubny1999, Gore2002}:
 \begin{eqnarray}
    &&   E_p(x^p) = e^x/p + \delta(x,p),    \label{eq.int09} \\
   &&    \delta(x,p) = -\frac{\sin(p\pi)}{ \pi } \int_{0}^{+\infty} 
                               \frac{ y^{p-1} e^{-xy} } { y^{2p} - 2y^{p}\cos(p\pi) + 1 } dy.    \label{eq.int10}  
 \end{eqnarray}
Function $\delta(p,x)$ for $0<p<2$ and $0\le x$ satisfies the relation, $|\delta(p,x)| \le |\delta(p,0)| = | p-1|/p$.
%

\section{\label{sec2}Analysis of the COBE monople data by Eqs.(\ref{eq.int09}) and (\ref{eq.int10})}

\begin{figure}
  \begin{center}
   \begin{tabular}{cc}
  \begin{minipage}[b]{0.45\linewidth}
    \includegraphics[width=\linewidth,clip]{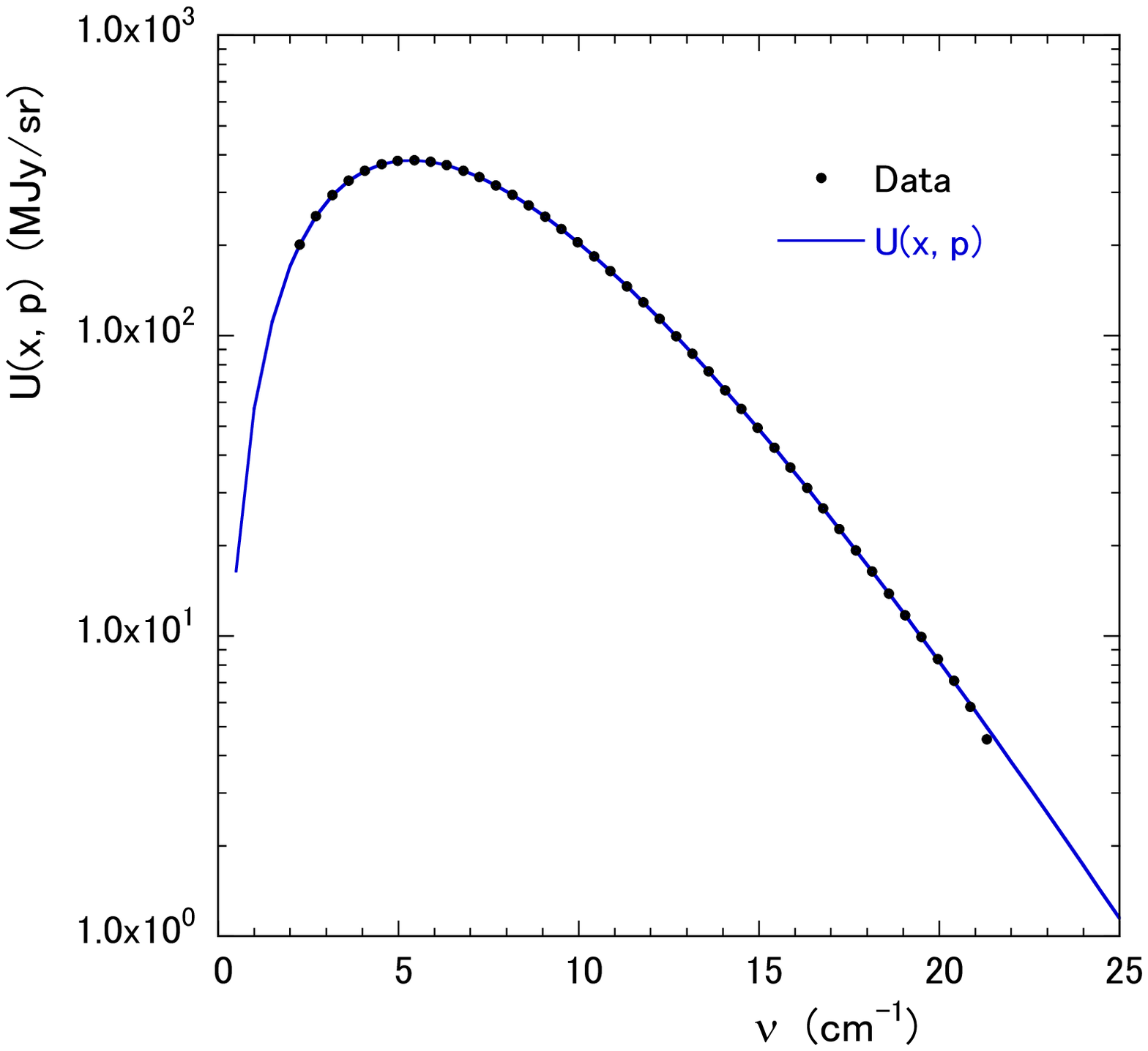}
    \caption{ \label{fig.firas02} Analysis of the COBE monopole data by Eq.~(\ref{eq.ana01}).   $x=0.528\nu$. }
  \end{minipage}
  \hspace{5mm}
  \begin{minipage}[b]{0.45\linewidth}
  \includegraphics[width=\linewidth,clip]{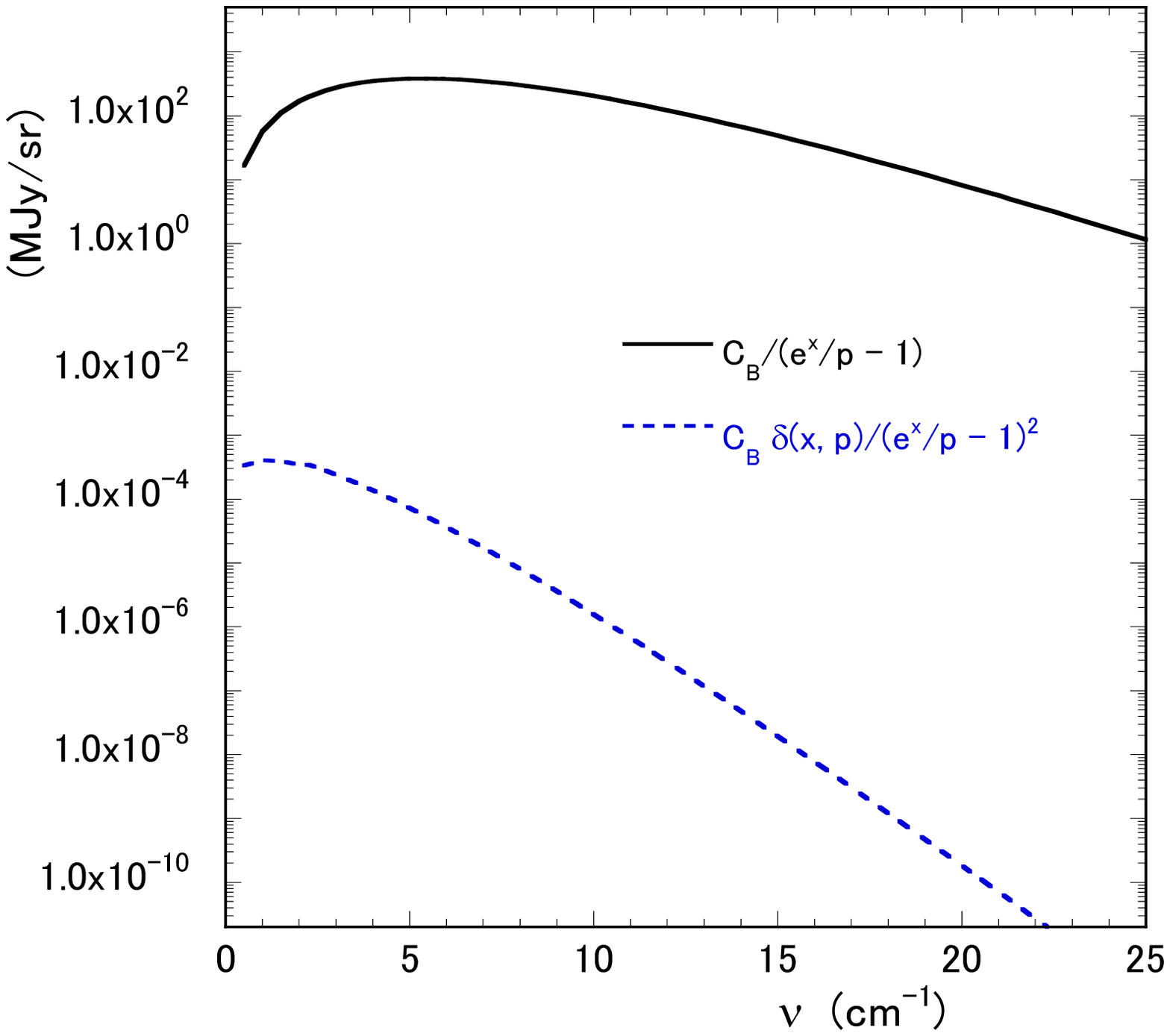}
  \caption{\label{fig.firas03} Contribution of first and second terms  
     on the right hand side of  Eq.~(\ref{eq.ana02}).  }
  \end{minipage} 
  \end{tabular}
 \end{center} 
\end{figure}

By the use of the integral representation of the ML function, Eq.~(\ref{eq.int09}), Eq.~(\ref{eq.int08}) is 
written as 
\begin{eqnarray}
     U(x,p) = {C_B}/( e^x/p - 1 + \delta(x,p) ),  \label{eq.ana01}
\end{eqnarray}
with two parameters, $T$ and $p$.
At first we analyze the COBE monopole data using Eq.(\ref{eq.ana01}). 
 The results are shown in Table \ref{tab1} and in  Fig.~\ref{fig.firas02}. 
As is seen from Table \ref{tab1}, conditions that $|p-1|<<1$ and $|\delta(p,x)|\le |p-1|/p<<1$ are satisfied.  
Then, we can expand Eq.~(\ref{eq.ana01}) as 
 \begin{eqnarray}
     U(x,p) = {C_B}/( e^x/p - 1) - {C_B \, \delta(x,p)}/(e^x/p - 1)^2.     \label{eq.ana02}
 \end{eqnarray}
>From the Analysis with Eq.(\ref{eq.ana02}), we obtain the same results with those in Table \ref{tab1}. 
Contribution from the first and second terms on the right hand side of Eq.~(\ref{eq.ana02}) 
with parameter values in Table \ref{tab1} are shown in Fig.~\ref{fig.firas03}.

 
%
\begin{table} [h]
  \caption{ \label{tab1} Analysis of the NASA COBE monopole data by Eq.(\ref{eq.ana01}).}
  \begin{center}
    \begin{tabular}{ccc}
     \br 
        $T$ (K) &  $(p-1)$ & $\chi^2$/N F \\
     \mr
        $2.72501 \pm 3\times 10^{-5}$ & $(1.1 \pm 3.5)\times 10^{-5}$ & 45.0/41\\
     \br
   \end{tabular}
  \end{center}
\end{table}

As the ratio of the second term to the first term on the right hand side of  Eq.~(\ref{eq.ana02}) 
becomes $\delta(x,p)/(e^x/p-1) <2 \times 10^{-6}$ over the range of the COBE monopole data, $1.20 \le x \le 11.26$,    
 we can approximate Eq.~(\ref{eq.ana02})  as,
 \begin{eqnarray} 
     U(x,p) \simeq  {C_B}/(e^x/p - 1) =  {C_B}/(e^{x-\ln p} - 1 ).    
   \label{eq.ana03} 
 \end{eqnarray}
Comparing Eq.(\ref{eq.int01}) and Eq.(\ref{eq.ana03}), we obtain an 
analytic relation,  $\mu = -\ln p$.

\section{\label{sec3}Concluding remarks}

1) If the Caputo fractional derivative is introduced into 
Eq.~(\ref{eq.int05}), 
contrary to the case of Riemann-Liouville fractional derivative~\cite{Biya2015}, 
we have fractional Bose-Einstein, Fermi-Dirac and Maxwell-Boltzmann distributions, 
where function ${\rm e}^x$ is replaced by the ML function, $E_p(x^p)$.  
\medskip\\
\noindent
2) Under the condition that $|p-1|<<1$, we can show that the analytic 
relation, $\mu = -\ln p$, is satisfied.  
In other words,  the fractional parameter $p$, where 
a kind of memory effect of the expanding universe would be included,  
has a role of inverse fugacity to the dimensionless 
chemical potential $\mu$.
\medskip\\
\noindent
3) Extension of statistical distributions has already been investigated from the non-extensive statistical approach~\cite{Tsallis2009}, where parameter $q$ is included. We would like to study how fractional parameter $p$ is related to $q$ and other approaches.
\\

\appendix

\section{\label{apdx1}Application of the Caputo fractional derivative to Eq.~(\ref{eq.int06})}

The Caputo fractional derivative~\cite{{Caputo1967},{Podlubny1999}} of  function $f(x)$ 
for $m=1,2,\ldots$ is defined as
 \begin{eqnarray}
   {}^C_0\! D^p_x f(x) =  \frac{1}{\Gamma(m-p)} \int_0^x (x-\tau)^{m-p-1} f^{(m)}(\tau) d\tau,
                                         \quad    m-1 < p < m,     \label{eq.cpd01} 
 \end{eqnarray} 
where $f^{(m)}(\tau)={d^mf(\tau)}/{d\tau^m}$. 
%
We consider the following equation,
 \begin{eqnarray}
      {}^C_0\!D^p_x R(x) =  aR(x) + b.    \label{eq.cpd02} 
 \end{eqnarray}
The Laplace transform of function $R(x)$ is defined as,  
 $\displaystyle{   \tilde{R}(s) = \mathcal{L}[ R(x); s ] = \int_0^\infty {\rm e}^{-sx} R(x) dx }$.
%
%
%
%
Applying the Laplace transform to Eq.~(\ref{eq.cpd02}), we obtain the following equation, 
 \begin{eqnarray}
    \tilde{R}(s) = b/\{s(s^p - a)\} + \sum_{k=0}^{m-1} R^{(m-k-1)}(0) 
s^{k-\nu}/(s^p-a). \label{eq.cpd03} 
 \end{eqnarray}
Using  the formula~\cite{Podlubny1999},
{\small
 \begin{eqnarray}
     &&\mathcal{L}[ x^{\beta-1} E_{\alpha,\beta}(\gamma x^\alpha);s ] 
            = s^{\alpha-\beta}/(s^\alpha - \gamma),   \quad  Re(s) > 
|a|^{1/\alpha},    \label{eq.cpd04}   \\
%
%
   && E_{\alpha,\beta}(z) = 
       \sum_{k=0}^\infty z^k/\Gamma(\alpha k + \beta),  
                         \quad \alpha > 0, \,\,\, \beta > 0,      \label{eq.cpd05}
 \end{eqnarray}
}
where $E_{\alpha,\beta}(z)$ is the two parameter ML function,  
we have 
{\small
 \begin{eqnarray}
    R(x) = b \{ E_p(ax^p) - 1 \} + R(0) E_p(ax^p)+ \sum_{k=1}^{m-1} 
R^{(m-k)}(0) x^{m-k-1} E_{p,m-k+1}(ax^p).    
      \label{eq.cpd07}
 \end{eqnarray}
}
Solutions $R(x)$ according to the values of $a$, $b$, and the initial 
conditions are shown in Table~\ref{tab2}.
 {\small
\begin{table} [h]
  \caption{\label{tab2}Solutions of  Eq.(\ref{eq.cpd02}).}
  \begin{center}
    \begin{tabular}{cccc}
     \br
        $(a, b)$  &          Initial conditions                       
            &     $R(x)$      & $f(x)=1/R(x)$ \\
     \mr
        $(1, 1)$  & $R(0)=\cdots =R^{(m-1)}(0)=0$                     
   & $E_p(x^p)-1$  & $1/( E_p(x^p)-1 )$ \\
     \mr
        $(1, -1)$ & $R(0)=2$, $R^{(1)}(0)=\cdots =R^{(m-1)}(0)=0$   & 
$E_p(x^p)+1$  & $1/( E_p(x^p)+1)$ \\
     \mr
        $(1, 0)$  &  $R(0)=1$, $R^{(1)}(0)=\cdots =R^{(m-1)}(0)=0$   &
 $E_p(x^p)$  & $1/ E_p(x^p) $ \\
     \br
   \end{tabular}
  \end{center}
\end{table}
}

\vspace{-4mm}
\section*{References}

\end{document}